\documentclass[sigconf,natbib=true,anonymous=false]{acmart}

\AtBeginDocument{%
  \providecommand\BibTeX{{%
    \normalfont B\kern-0.5em{\scshape i\kern-0.25em b}\kern-0.8em\TeX}}}
\usepackage{pifont}
\usepackage{caption}
\usepackage{subcaption}
\usepackage{booktabs}       
\usepackage{bbm}
\usepackage{amsmath}
\usepackage{tablefootnote}
\usepackage{threeparttable}
\usepackage{balance}
\definecolor{cmarkcolor}{RGB}{21, 164, 64}
\newcommand{\stitle}[1]{\noindent\textup{\textbf{#1}}}
\newcommand{\cmark}{\color{cmarkcolor}
\ding{51}}%
\definecolor{xmarkcolor}{RGB}{177, 0, 4}
\newcommand{\xmark}{\color{xmarkcolor}\ding{55}}%
\newcommand{\ourdataset}{$\mathsf{MobileRec}$}

\acmYear{2023}
\copyrightyear{2023}
\acmSubmissionID{6396-resouce paper}
\acmConference[SIGIR'23]{Proceedings of the 46th ACM SIGIR Conference on Research and Development in Information Retrieval,}{July 23-27, 2023}{Taipei, Taiwan}
\newcommand{\myvalue}[1] {$\mathtt{#1}$}

\begin{document}

\title{
MobileRec: A Large-Scale Dataset for Mobile Apps Recommendation
}

\author{M.H. Maqbool}
\affiliation{%
  \institution{University of Central Florida}
  \city{Orlando}
  \state{FL}
  \country{USA}
}
\email{hasanmaqbool@knights.ucf.edu}

\author{Umar Farooq}
\affiliation{%
  \institution{Independent Researcher}
  \city{Mountain View}
  \state{CA}
  \country{USA}
}
\email{ufarooq.cs@gmail.com}

\author{Adib Mosharrof}
\affiliation{%
  \institution{University of Kentucky}
  \city{Lexington}
  \state{KY}
  \country{USA}
}
\email{amo304@g.uky.edu}

\author{A.B. Siddique}
\affiliation{%
  \institution{University of Kentucky}
  \city{Lexington}
  \state{KY}
  \country{USA}
}
\email{siddique@cs.uky.edu}

\author{Hassan Foroosh}
\affiliation{%
  \institution{University of Central Florida}
  \city{Orlando}
  \state{FL}
  \country{USA}
}
\email{hassan.foroosh@ucf.edu}


\begin{abstract}
Recommender systems have become ubiquitous in our digital lives, from recommending products on e-commerce websites to suggesting movies and music on streaming platforms. 
Existing recommendation datasets, such as Amazon Product Reviews and MovieLens, greatly facilitated the research and development of recommender systems in their respective domains.
While the number of mobile users and applications (aka apps) has increased exponentially over the past decade, research in mobile app recommender systems has been significantly constrained, primarily due to the lack of high-quality benchmark datasets, as opposed to recommendations for products, movies, and news.
To facilitate research for app recommendation systems, we introduce a large-scale dataset, called {\ourdataset}.
We constructed {\ourdataset} from users’ activity on the Google play store. 
{\ourdataset} contains $19.3$ million user interactions (i.e., user reviews on apps) with over $10$K unique apps across $48$ categories.
{\ourdataset} records the sequential activity of a total of $0.7$ million distinct users. 
Each of these users has interacted with no fewer than five distinct apps, which stands in contrast to previous datasets on mobile apps that recorded only a single interaction per user.
Furthermore, {\ourdataset} presents users' ratings as well as sentiments on installed apps, and each app contains rich metadata such as app name, category, description, and overall rating, among others. 
We demonstrate that {\ourdataset} can serve as an excellent testbed for app recommendation through a comparative study of several state-of-the-art recommendation approaches.
The quantitative results can act as a baseline for other researchers to compare their results against.
The {\ourdataset} dataset is available at 
\textcolor{blue}{\url{https://huggingface.co/datasets/recmeapp/mobilerec}.}

\end{abstract}


\maketitle
\section{Introduction}
\label{sec:intro}
\begin{figure}[t!]
    \centering
    \includegraphics[width=\linewidth]{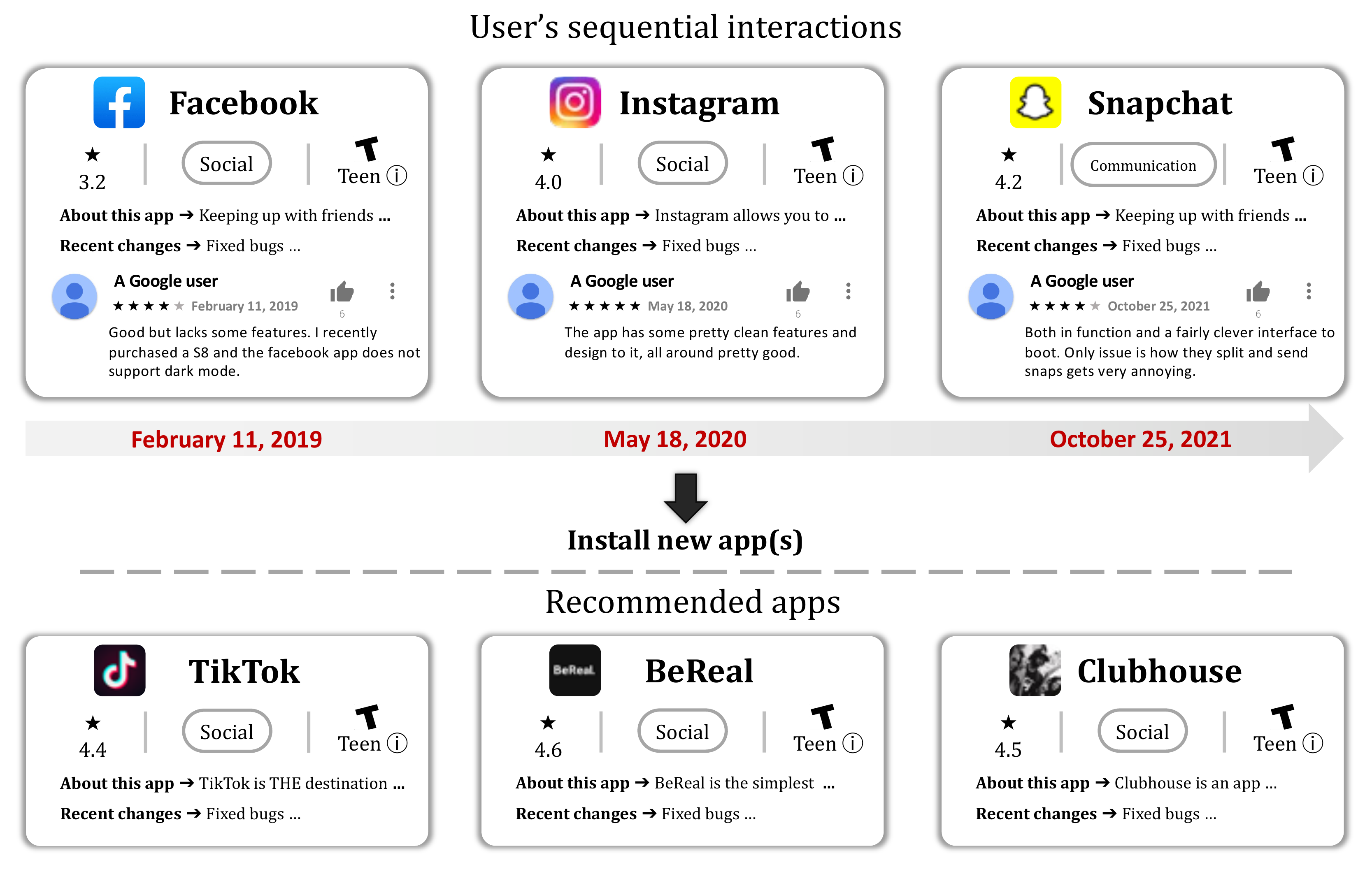}
    \caption{An Example of a sequence of the user activity. Based on past user interactions (e.g., app installations), the app recommendation system recommends new apps to install.}
    \label{fig:app-recommndation-exampple}
    \vspace{-15pt}
\end{figure}

Mobile apps have seen exponential growth in the last decade and over $5$ billion users~\cite{number-of-smartphone-users} utilize them for a variety of reasons, including social media, entertainment, news, productivity, and ride-sharing, among others. 
As a result of this boom, Google Play~\cite{googleplay} and Apple App store~\cite{appstore} host more than $3.5$ and $2.2$ million apps, respectively~\cite{number-of-apps-on-stores}.
The increasingly crowded app marketplaces pose a significant challenge for users to discover apps that align with their preferences effectively.
Personalized app recommendations can relieve users' cognitive overload and improve the app installation experience.
As illustrated in Figure~\ref{fig:app-recommndation-exampple}, an app recommendation system has the capability to suggest new applications to users based on their previous app installations and interactions.
Although Google Play and the App Store employ app recommendation techniques for suggesting apps to their users potentially leveraging user data collected internally, the research in app recommendation is almost nonexistent.

\begin{table*}[h!]
\centering
\caption{Comparison of existing mobile apps datasets with MobileRec.}
    \begin{tabular}{lccccccc}
       \toprule
       Dataset features & Top 20 Apps~\cite{top20-dataset} & {\sc RRGen}~\cite{rrgen} & {\sc AARSynth}~\cite{aarsynth} & Srisopha et al.~\cite{srisopha-how}$^\ast$ & {\sc PPrior}~\cite{pprior}$^\dagger$  & {\ourdataset}  \\
       \midrule
       Number of Reviews & 200K & 309K & 2.1M & 9.3M & 2.1M & 19.3M \\
       Number of Apps & 20 & 58 & 103 & 1,600 & 9,869 & 10,173 \\
       Number of App Categories & 9 & 15 & 23 & 32 & 48 & 48 \\
       \midrule
       Multiple Reviews by a Single User & \xmark & \xmark & \xmark & \xmark & \xmark & \cmark \\
       App Metadata  & \xmark & \xmark & \cmark & \xmark & \xmark & \cmark \\
       Reviews Rating & \cmark & \cmark & \cmark & \cmark & \cmark & \cmark \\
       Review Text & \cmark & \cmark & \cmark & \cmark & \cmark & \cmark \\
       Review Timestamp & \xmark & \xmark & \xmark & \cmark & \cmark & \cmark \\
       \bottomrule
       \multicolumn{4}{l}{
       $^\ast$ ~\cite{srisopha-how} is not publicly available.
       $^\dagger$ ~\cite{pprior} contains only negative user reviews.}  
    \end{tabular}
    \vspace{-5pt}
    \label{tab:compare-mobile-datasets}
\end{table*}

Recommendation systems have demonstrated remarkable effectiveness in a wide range of domains, such as news~\cite{wu2020mind}, movies~\cite{diao2014jointly}, products~\cite{yan22personalized,tanjim20attentive}, fashion~\cite{iwata2011fashion}, fitness~\cite{ni19modeling}, toys, beauty, CDs~\cite{mcauley2015image, he2016ups, ni2019justifying}, to name a few~\cite{rappaz21recommendation}.
The availability of datasets for specific domains that contain user-item interactions~\cite{amazon-dataset, koren2008factorization, movielens} have played a critical role in the development and improvement of recommendation systems.
In earlier research items and users were typically represented using their unique identifiers, and their respective interactions, such as rating scores, were used to learn useful relationships that could aid in recommendation tasks. These relationships were typically learned using collaborative filtering techniques or similar methods~\cite{koren2008factorization}.
Recently, incorporating sequential user interactions into the development of recommendation systems has led to significant improvements in their performance~\cite{kang2018self}. 
By taking into account the sequential patterns of user interactions with items~\cite{ni2019justifying} (e.g., the order in which they viewed or purchased items), these methods can better capture the user's preferences and provide more accurate recommendations~\cite{kang2018self}. 
Furthermore, the development of novel techniques such as recurrent neural networks (RNNs)~\cite{elman1990finding,hochreiter1997long} and attention-based models~\cite{vaswani2017attention} that effectively handle sequential data has shown exceptional performance in recommendation tasks~\cite{yan22personalized,tanjim20attentive}.

There are several notable datasets that focus on mobile applications, such as the {\sc Top 20 apps} dataset~\cite{top20-dataset}, {\sc RRGen}~\cite{rrgen}, {\sc AARSyth}~\cite{aarsynth}, Srisopha et al.\cite{srisopha-how}, and {\sc PPrior}\cite{pprior}, among others.
However, {\sc Top 20 Apps} and {\sc RRGen} only contain a limited number of reviews, typically in the hundreds of thousands. Moreover, they only cover fewer than 100 apps from less than 20 categories.
Despite providing larger datasets with millions of reviews from thousands of apps, datasets such as {\sc AARSynth}, Srisopha et al., and {\sc PPrior} are not amenable to building effective app recommendation systems, since they lack unique user identifiers. 
Furthermore, the dataset from Srisopha et al.~\cite{srisopha-how} is not publicly available and {\sc PPrior} only provides negative user reviews.
A comparison of mobile app datasets is presented in Table~\ref{tab:compare-mobile-datasets}.
This work attempts to fill this research gap by providing a large-scale, rich, and diverse
benchmark dataset, we call {\ourdataset}, to facilitate researchers in developing app recommendation systems. 
To the best of our knowledge, \emph{this is the only mobile app dataset that offers sequential and multiple interactions per user} with apps.
We also present a comparison of our proposed dataset with the latest versions of the well-established recommendation datasets in various domains in Table~\ref{tab:compare-rec-datasets}.

{\ourdataset} is a large-scale dataset constructed from users' activity on the Google Play Store, containing $19.3$ million user interactions (i.e., user reviews along with rating) across $10$K unique apps in $48$ categories. 
It records the sequential activity of $0.7$ million unique users, each of whom interacted with at least five apps.
Every user interaction within our dataset consists of essential information, such as the user's rating, textual review, and the review date, pertaining to installed apps.
Moreover, every individual app in {\ourdataset} is equipped with abundant metadata, such as the app's title, category, long-form textual description, overall average rating, developer information, content rating, and additional pertinent details.
Table~\ref{tbl:feature_desc} presents important features of the dataset.
To stimulate research in app recommendation systems, {\ourdataset} is on par with well-established and widely recognized datasets from various other domains.
On top of app recommendation, our comprehensive dataset can offer significant insights and serve as a valuable resource for diverse research and data analysis pursuits, such as sentiment analysis, app development and optimization, market research, and fraud detection, among others.

\begin{table*}[t!]
\centering
\caption{{\ourdataset}'s comparison with the latest versions of well-known recommendation datasets in different domains. We used the Amazon Reviews dataset from 2018, Yelp's 2022 version, and the latest release of ML-25M from 12/2019 to produce these statistics.}
\begin{tabular}{lcccc}
\toprule
\multicolumn{1}{c}{\textbf{Datasets}}                                       & \textbf{Amazon Reviews} & \textbf{Yelp} & \textbf{ML-25M} & \textbf{MobileRec (Ours)} \\ \hline
\multicolumn{1}{l|}{Total interactions (in millions)}                                     & 233.1               & 6.99       & 25.0        & 19.3                  \\
\multicolumn{1}{l|}{Users with at least five interactions (in millions)}  & 10.6                & 0.29        & 0.16          & 0.70                    \\
\multicolumn{1}{l|}{Items with at least 15 interactions (in thousands)} & 2072                 & 77.58         & 20.59           & 10.17                     \\
\multicolumn{1}{l|}{Maximum number of interactions by a single user}                          & 446                     & 3048          & 32202           & 256                       \\
\multicolumn{1}{l|}{Minimum number of interactions by a single user}                          & 1                       & 1             & 20              & 5                         \\
\multicolumn{1}{l|}{Average number of interactions per user}                            & 5.32                  & 3.52        & 153.81         & 27.56                     \\
\multicolumn{1}{l|}{Maximum number of interactions on a single item}                           & 13,560                   & 7,673          & 81,491           & 14,345                     \\
\multicolumn{1}{l|}{Minimum number of interactions on a single item}                           & 1                       & 5             & 1               & 20                        \\
\multicolumn{1}{l|}{Average number of interactions per item}                             & 15.45                 & 46.49       & 423.39         & 1896.88                   \\ 
\bottomrule
\end{tabular}
\vspace{-10pt}
\label{tab:compare-rec-datasets}
\end{table*}

We also show that {\ourdataset} can serve as a valuable experimental testbed for app recommendation research through a comparative evaluation of several cutting-edge recommendation techniques.
Specifically, we employed several existing general and sequential recommendation systems, including Pop (i.e., the popularity of items), SASRec~\cite{kang2018self}, ELECRec~\cite{chen2022elecrec}, BERT4Rec~\cite{sun2019bert4rec}, HGN~\cite{ma2019hierarchical}, SINE~\cite{tan2021sparse}, LightSANs~\cite{fan2021lighter}, GRU4Rec~\cite{tan2016improved}, and GCSAN~\cite{xu2019graph} using {\ourdataset}.
Our analysis of the results obtained by these methods on the {\ourdataset} has revealed some intriguing trends. 
Notably, we observed that ELECRec~\cite{chen2022elecrec} had performed significantly better than GRU4Rec~\cite{tan2016improved} on the \myvalue{Amazon}-\myvalue{Beauty} dataset, achieving over $300\%$ improvement in the \myvalue{Hit@5} metric. 
However, when tested on {\ourdataset}, ELECRec exhibited a considerable drop in performance compared to GRU4Rec on the same metric.
Similarly, we observe that LightSANs~\cite{fan2021lighter} had achieved a $3.91\%$ and $2.65\%$ increase in \myvalue{Hit@10} and \myvalue{NDCG@10} metrics, respectively, compared to SASRec~\cite{kang2018self}. 
However, on {\ourdataset}, SASRec~\cite{kang2018self} outperforms LightSANs~\cite{fan2021lighter}.
This signifies the inherent dynamism and fleeting nature of user-item interactions and user interests in the {\ourdataset} dataset, which gives rise to complex user interaction history.
This observation has important implications for state-of-the-art recommendation models.
The numerical results presented in this work can also serve as a baseline for fellow researchers to evaluate their own results and objectively assess the quality and effectiveness of different approaches.
Moving forward, numerous natural language processing techniques, such as advanced text representation methods and pre-trained language models~\cite{devlin2018bert,radford2019language}, have the potential to unlock new research avenues for app recommendations using {\ourdataset}. 
These techniques hold great promise for enhancing the quality and effectiveness of app recommendation systems, and as such, are of particular interest to the research community.

Specifically, this work makes the following contributions:
\begin{itemize}
\item We present {\ourdataset}, the most extensive collection of sequential user-app interactions to date, comprising over $19$ million interactions across a diverse range of over $10$ thousand distinct apps from Google Play, spanning all categories. Notably, this is the only mobile app dataset that features multiple interactions per user. 
\item Our experimental study investigates the dynamics of user-app interactions and demonstrates the practical utility of {\ourdataset} by employing several state-of-the-art recommendation systems and establishing baseline results for the research community.
\end{itemize}

\section{Related Work}
\label{sec:related}


\subsection{App Datasets and Recommendation}
There are several existing datasets for user reviews of mobile apps as listed in Table~\ref{tab:compare-mobile-datasets}. 
In an early effort to collect user reviews, Iacob and Harrison~\cite{iacob2013retrieving} collected 3,279 reviews for 161 mobile apps and analyzed feature requests from users. 
Khalid et al.~\cite{khalid2014mobile} and ~\cite{khalid2013identifying} focused on iOS apps and prepared a dataset of 6,390 user reviews for 20 apps. McIlroy et al.~\cite{mcilroy2016analyzing} used 601,221 user reviews for 12,000 mobile apps to study negative reviews on app stores. Maalej and Nabil~\cite{maalej2015bug} collected a much larger dataset with 1.3 million reviews for 1,186 apps. These datasets focus on user complaints and user-developer dialogue understanding. Moreover, these datasets are not publicly available. 

Top 20 Apps~\cite{top20-dataset} is available on Kaggle, which contains 200 thousand reviews for 20 apps spanning 9 categories. 
This dataset provides rating scores and text for the reviews. 
{\sc RRGen}~\cite{rrgen} contains more than 309 thousand reviews from 58 apps. Similar to the Top 20 Apps, {\sc RRGen} provides only rating scores and text of reviews. Both datasets do not provide app metadata, the timestamp of review, and a unique identifier for the user. 

{\sc AARSynth}~\cite{aarsynth} provides more than two million user reviews spanning over a hundred apps, including app metadata. 
Reviews of this dataset also miss out on key information similar to the above-mentioned datasets. 
The dataset by Srisopha et al.~\cite{srisopha-how} includes over 9 million user reviews from 1,600 apps. 
This dataset has review timestamps, which can help to understand reviews in the context of the time period. However, this dataset does not include the user's unique identifier and app metadata. 
Please note that dataset by Srisopha et al.~\cite{srisopha-how} is not publicly available. 

More recently, the {\sc PPrior}~\cite{pprior} dataset provided more than 2 million reviews for over 9 thousand apps spanning $48$ categories from Google play. 
This dataset provides rating scores, review text, and timestamps of reviews. 
Nonetheless, user identifier on interactions (i.e., reviews) and app metadata is not provided. 
Furthermore, {\sc PPrior} only provided negative user reviews (i.e., reviews with ratings 1 and 2 only).
Hence, making {\sc PPrior} not suitable for building robust app recommendation systems. 
In this work, we provide a larger dataset, called {\ourdataset}, than all of the above-mentioned datasets for mobile apps, including each user's unique identifier and timestamps with interactions.
Furthermore, each user has a least five interactions with apps and all the included apps have at least 15 interactions, which makes it an ideal testbed for mobile recommendation systems.

\begin{table*}[t]
\caption{Description of the various important features in the {\ourdataset} dataset. 
}
\begin{tabular}{cl}
\toprule
\textbf{Feature} & \multicolumn{1}{c}{\textbf{Description}}                                                                                                                                                                  \\ \hline
uid              & \begin{tabular}[c]{@{}l@{}}16-characters alphanumeric uid represents a user uniquely. It also anonymizes the user. \\ Example: Aj0Sm6myfh6YN3Rn, 3pZUhksFIcLjEXtl, dvx0dqXTKtHUmY3O\end{tabular}          \\ \hline
app\_package     & \begin{tabular}[c]{@{}l@{}}Represents android package name of an application and uniquely identifies an application.\\ Example: com.google.android.calculator, com.king.crash, org.wikipedia\end{tabular} \\  \hline
app\_name     & \begin{tabular}[c]{@{}l@{}}Title of the app as displayed on Google Play.\\ Example: Candy Crush Saga, MONOPOLY - Classic Board Game\end{tabular} \\
\hline
app\_category    & \begin{tabular}[c]{@{}l@{}} The category of the app. \\
Example: Entertainment, Finance, Productivity
\end{tabular}         \\
\hline
review           & Text review given by a user to an application                                                                                                                                                             \\ \hline
rating           & \begin{tabular}[c]{@{}l@{}}Numeric rating the user has given to an application.\\ Example: 5, 2, 1, 4                 \end{tabular}                                                                                                                 \\ \hline
votes            &    \begin{tabular}[c]{@{}l@{}}The number of users who found this review helpful.\\ Example: 1, $\cdots$, 6,  $\cdots$                \end{tabular}                                                                                                                                                                                           \\ \hline
date             &         
\begin{tabular}[c]{@{}l@{}}The date of the specific user/item interaction, i.e., review date. \\ Example: October 21, 2018, November 4, 2021, January 16, 2021                                        \end{tabular} \\ \hline

                                                         formated\_date &\begin{tabular}[c]{@{}l@{}}The date of the specific user/item interaction. (review date in YYYY-MM-DD format). \\ Example: 2018-10-21, 2021-11-04, 2021-01-16                                        \end{tabular}                                                                                               \\ \hline
unix\_timestamp  & 

\begin{tabular}[c]{@{}l@{}}The review date converted into unix timestamp. \\ Example: 1.540094e+09	1.547788e+09, 1.610773e+09	 \end{tabular}                       
                                                                                                                                          \\
                                                                                                                                          \bottomrule       
\end{tabular}
\label{tbl:feature_desc}
\end{table*}

\subsection{Existing Recommendation Datasets}
Recommendation datasets are available across a diverse array of domains, encompassing everything from e-commerce platforms like Amazon, to the entertainment industry with datasets focused on movies, and even extending to online gaming platforms and beyond. 
These diverse datasets offer a wealth of opportunities for researchers to investigate and improve the effectiveness of recommendation systems across a broad range of use cases.

{\sc Amazon Product Reviews}~\cite{mcauley2015image, he2016ups, ni2019justifying} is a large-scale dataset that contains 233.1 million reviews in the updated version of 2018. 
There are $29$ categories in the {\sc Amazon Product Reviews} dataset. 
Note that, an earlier version of this dataset which was released in 2014 had 142.8 million interactions and 24 categories. 
This dataset provides review text, rating, and helpfulness votes along with descriptions, category information, price, and brand, among others, as product metadata. 
{\sc Amazon Product Reviews} dataset contains several categories including \myvalue{All} \myvalue{Beauty}, \myvalue{Books}, \myvalue{CDs} \myvalue{and} \myvalue{Vinyl}.
Each of these categories contains thousands to millions of reviews. 
For example, \myvalue{All} \myvalue{Beauty} has $0.37$ million reviews and $32$K unique products, \myvalue{Books} has $51$ million reviews and $2$ million products, and \myvalue{CDs} \myvalue{and} \myvalue{Vinyl} contains $1$ million reviews and $544$K products. 
{5-core} is a small version of {\sc Amazon Product Reviews } dataset that contains at least 5 user interactions, which makes it useful for building robust recommendation systems. 
Since {\sc Amazon Product Reviews} is a very large dataset, several of the existing works\cite{chen2022elecrec, kang2018self} just employ some of the categories from the dataset such as \myvalue{Books}, \myvalue{Beauty}, \myvalue{CDs}, and \myvalue{Games} to benchmark their proposed methods. 

{\sc Yelp}~\cite{yelp-dataset} dataset has several releases starting from 2018 to 2022. 
The latest version, {\sc Yelp2022} has $1.9$ million unique users, $150$K distinct items, and $6.9$ million reviews. 
Similarly, {MovieLens}~\cite{movielens} offers {\sc ML-25M}~\cite{harper2015movielens}, which comes with $25$ million ratings and $1$ million tag applications across $62$K movies. 
The past versions of this dataset include {\sc ML-100k} and {\sc ML-1M}.

In addition to the large-scale datasets mentioned above, a wide variety of datasets from diverse domains have played a vital role in advancing research within their respective fields. 
In the following, we briefly discuss other recommendation datasets.
{\sc Steam} contains reviews and game information crawled from stream~\cite{kang2018self}. 
{\sc Book-Crossing}~\cite{ziegler2005improving} and {\sc GoodReads}~\cite{wan2018item, wan2019fine} provide user-item interactions where interaction type is rating.
%
{\sc DIGINETICA}\cite{recbole[1.0]} contains $0.2$ million users, $184$K items and about $1$ million interactions, where interaction type is user-clicks. 
This dataset has been compiled from user sessions extracted from e-commerce search engine logs. 
{\sc Twitch}~\cite{rappaz2021recommendation} offers $474$ million interactions by $15$ million users on $6$ million items, and the interaction type is user-click. 
These datasets have enabled the development of several recommendation systems~\cite{diao2014jointly,yan22personalized,tanjim20attentive,iwata2011fashion,ni19modeling,mcauley2015image, he2016ups, ni2019justifying,rappaz21recommendation} in their respective domains. 
We expect that {\ourdataset} plays a similar role to ignite research in building robust app  recommender systems.

\section{{\ourdataset} Dataset}
\label{sec:dataset}

\begin{figure*}[t!]
    \centering
    \begin{subfigure}[b]{0.23\textwidth}
        \centering
        \includegraphics[width=\linewidth]{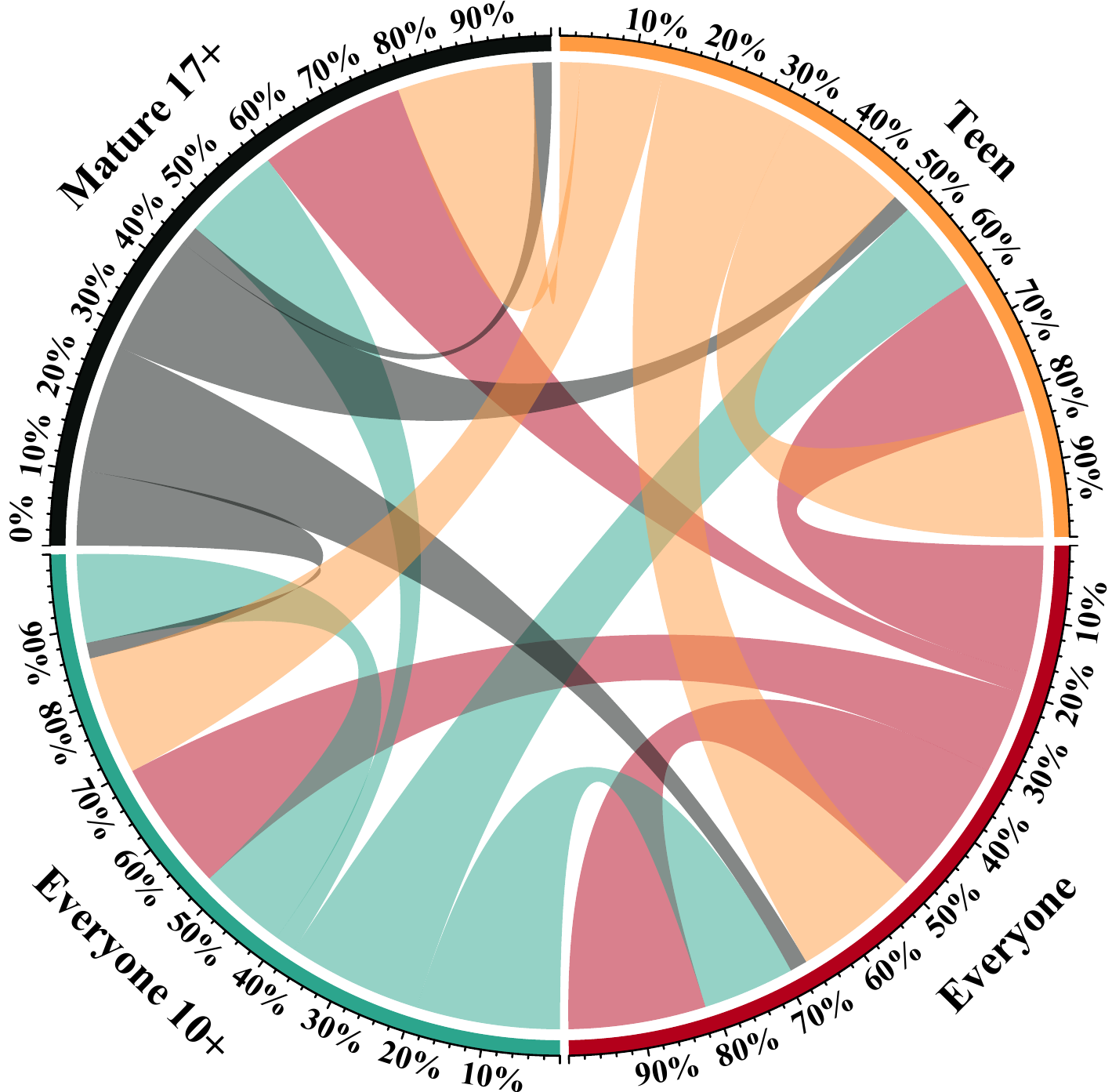}
        \caption{content\_rating based users' sequential interactions trend.}
        \label{aaa}
    \end{subfigure}
    \hfill
    \begin{subfigure}[b]{0.225\textwidth}
        \centering
        \includegraphics[width=\linewidth]{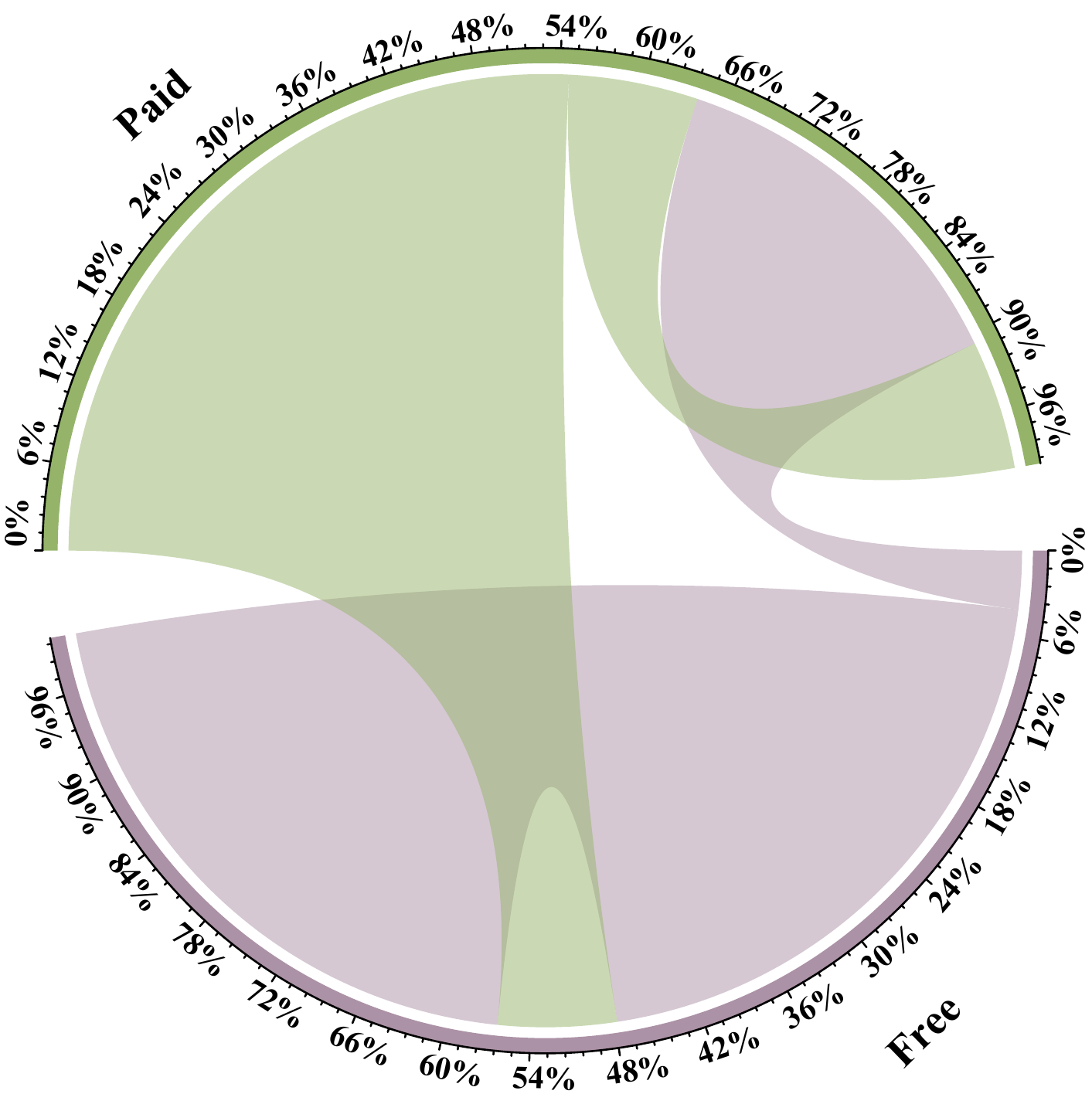}
        \caption{App price-based users' sequential interactions trend.}
        \label{bbb}
    \end{subfigure} 
    \hfill
    \begin{subfigure}[b]{0.23\textwidth}
        \centering
        \includegraphics[width=\linewidth]{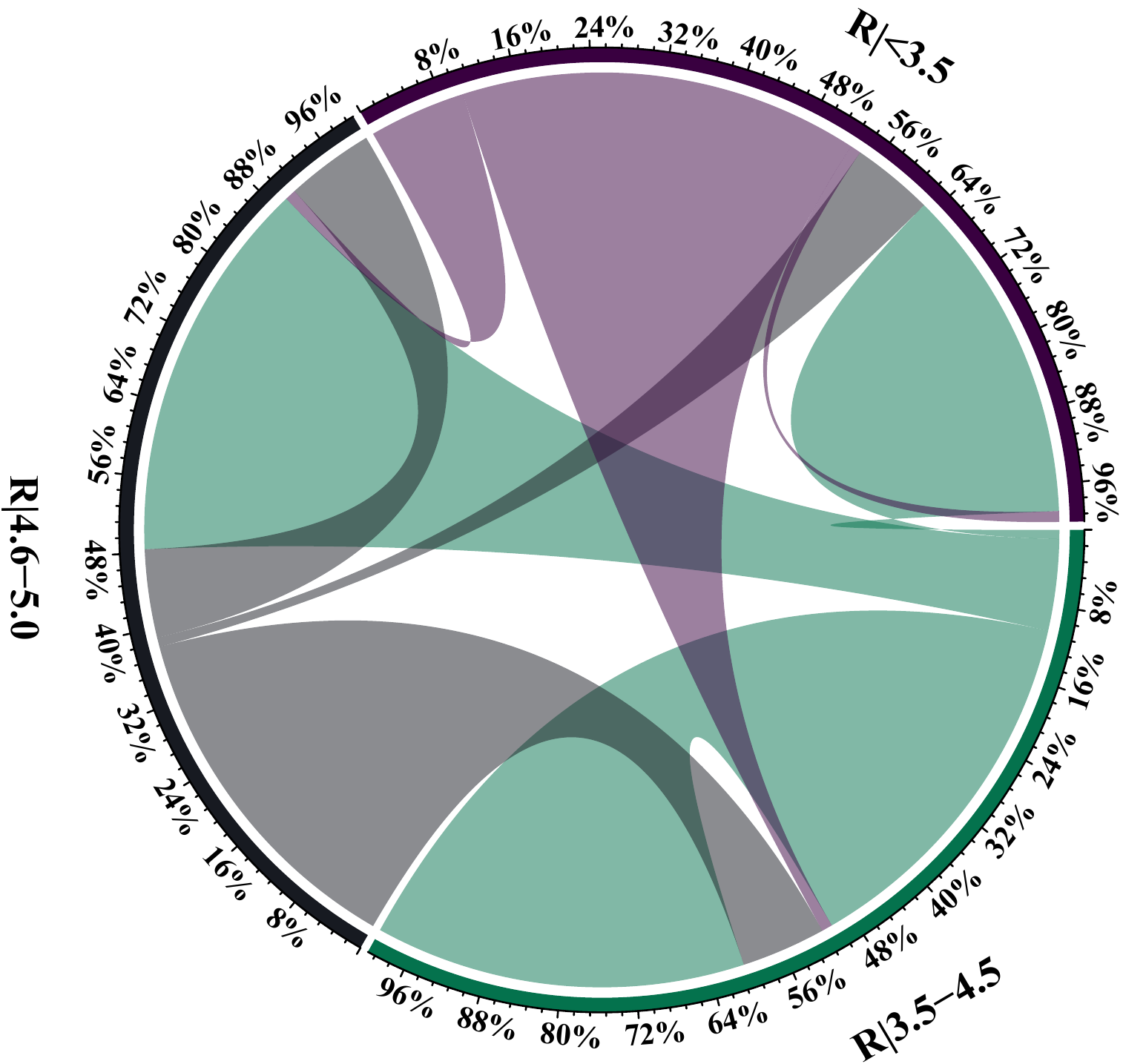}
        \caption{App rating based users' sequential interactions trend.}
        \label{ccc}
    \end{subfigure} 
    \hfill
    \begin{subfigure}[b]{0.23\textwidth}
        \centering
        \includegraphics[width=\linewidth]{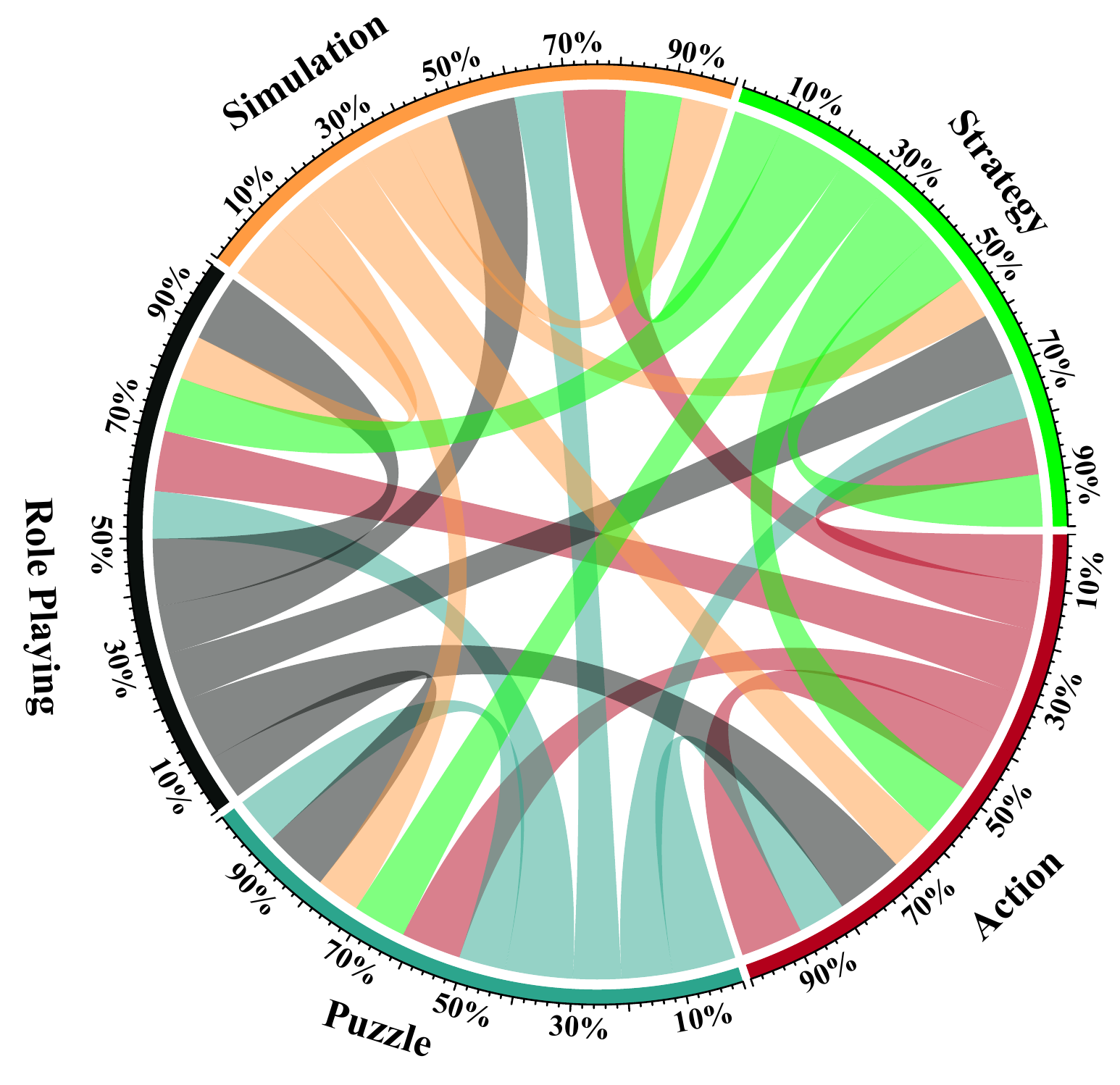}
        \caption{Users' sequential interactions trend within top-5 categories.}
        \label{ddd}
    \end{subfigure} 
    \caption{Users' sequential interactions trend with respect to conten\_rating, price tier, rating and within top-5 categories.}
\label{fig:migrations}
\end{figure*}

\subsection{Dataset Construction}
Google Play~\cite{googleplay} serves as the default app store for android users to install the apps and express their opinion about the apps. 
Consequently, it hosts an enormous amount of user interactions through user reviews.
The user reviews are dynamically loaded upon scrolling the web page. Therefore, traditional web scraping tools cannot facilitate downloading of large-scale user reviews data. 
To automate the scrolling of the page and perform other click events (e.g., longer reviews are not displayed fully by default), we used Selenium WebDriver~\cite{selenium}.

To obtain the data, the first step was to collect the package names (i.e., unique identifiers) of all the apps directly accessible from Google Play by navigating through all the app category pages as well as top charts. 
Then, we used this information to download app metadata. The metadata about an app included details such as the app name, developer, category, and textual description, among others. 
The user reviews associated with each app were extracted by accessing the user review section of the app, recursively scrolling the page, and extracting the text of the review along with the user's rating of the app, user's information (i.e., anonymized later on), and timestamp until no more reviews were available for the given app.
Table~\ref{tbl:feature_desc} presents important features of the dataset.
We used several filters and sorting mechanisms (e.g., newest, rating) to allow for downloading the maximum number of user reviews.

To ensure the quality of the data, several checks were put in place to eliminate duplicate entries, incorrect information, and other errors that may have occurred during the crawling process.
We removed the duplicates from the data before processing the data further for converting it to a 5-core. 
We also anonymized the users by introducing a 16-character alphanumeric \textit{uid}. 
We only kept those users who have more than 5 reviews. 
The items having less than 15 reviews were removed from the final dataset.
The end result was a large-scale 5-core dataset of user reviews (i.e., timestamped interactions) for apps available on Google Play, providing valuable insights into user opinions and experiences with various apps.
Table~\ref{tab:compare-mobile-datasets}~and~\ref{tab:compare-rec-datasets} provide important statistics about the dataset.

\subsection{Dataset Features}
We introduce various features (also presented in Table~\ref{tbl:feature_desc}) in the dataset which will be helpful in discussing the dataset trends. 
The \textit{uid} is the unique 16-character alphanumeric id of a user, which also serves as anonymization of the users, since {\ourdataset} does not provide the actual user IDs for privacy reasons. 
The \textit{review} is the actual text review by a user. 
The \textit{app\_package} is the unique android package name of an application. 
The \textit{rating} represents the actual rating given to an application (identified by {app\_package}) by a user. 
The \textit{formated\_date} is used for sorting the user interactions in chronological order. For further reference, Table \ref{tbl:feature_desc} can be consulted. 

\subsection{Dataset Analysis}
In this section, we present various trends in the {\ourdataset} dataset, e.g., how do users migrate from one category to the other with respect to their review-based interactions, how do users' behavior evolve with respect to application pricing, how do users migrate from application to application with respect to application's content rating (i.e., Mature 17+, Teen, Everyone 10+, and Everyone).

\begin{figure}[t!]
    \centering
    \includegraphics[width=0.35\textwidth]{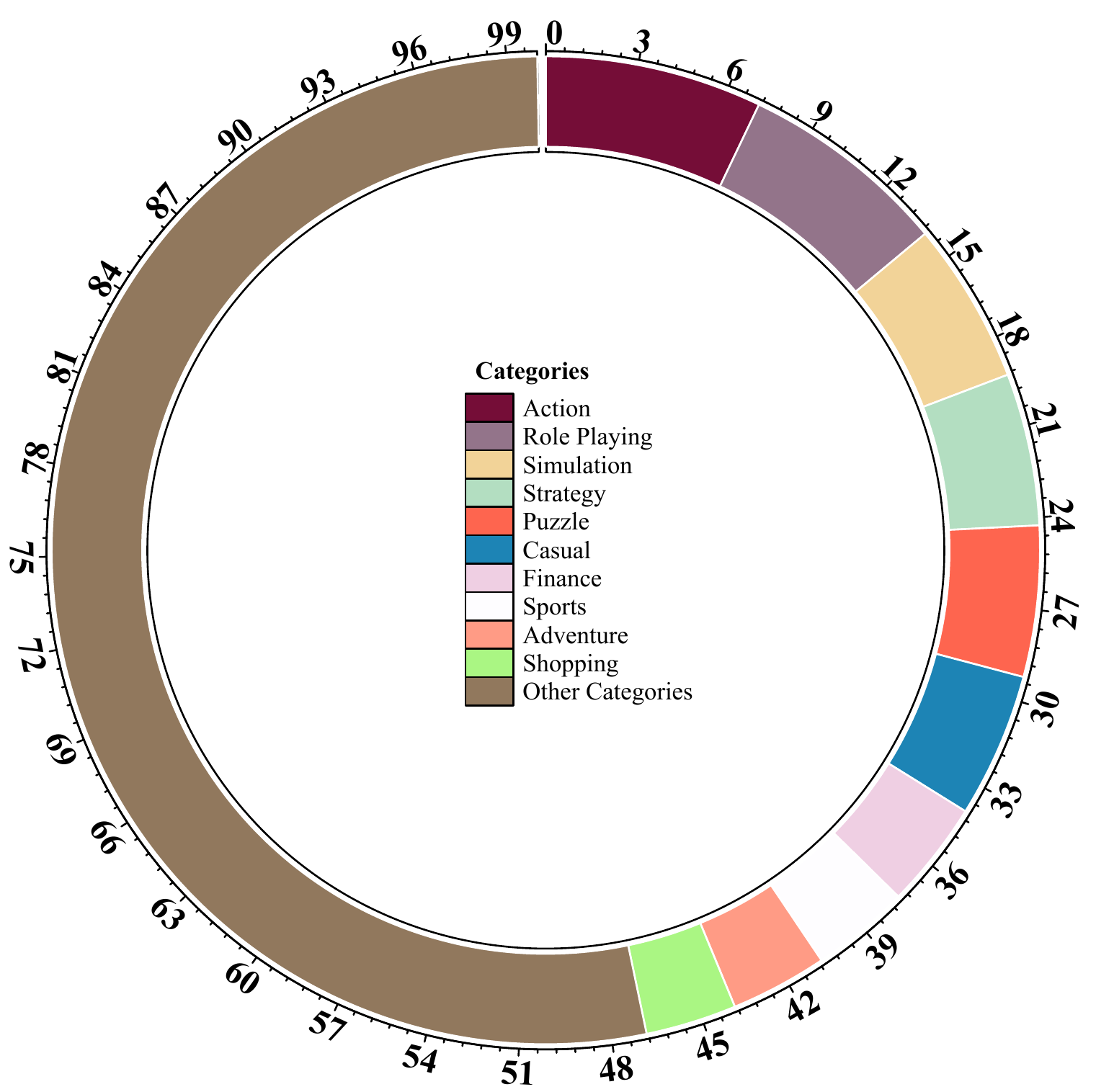}
    \captionsetup{justification=centering}
    \caption{Top-10 categories with respect to the number of reviews. The numbering around the dial depicts the percentage, e.g., apps belonging to the action category add up to $\approx7\%$ of total reviews in the {\ourdataset} dataset.}
    \label{fig:circular_bar_plot}
    \vspace{-15pt}
\end{figure}

We capture an interaction-based snapshot of users' dynamic behavior in Figure \ref{fig:migrations}(a), \ref{fig:migrations}(b), \ref{fig:migrations}(c) \& \ref{fig:migrations}(d). 
We consider the first two user interactions for capturing the migration behavior. 
For example, in Figure \ref{fig:migrations}(a), it can be noticed that the majority of users with their interaction in \myvalue{Mature 17+} categories at a given timestep $t$, ended up interacting with \myvalue{Everyone} and the \myvalue{Teen} content-rated apps in their interaction at timestep $t+1$. 
Similarly, an expected trend is revealed in Figure~\ref{fig:migrations}(b) which illustrates price-based sequential user interactions.
Figure~\ref{fig:migrations}(b) points out that very few users who interact with \myvalue{free} apps at timestep $t$, interact with \myvalue{paid} apps at their timestep $t+1$. 
Conversely, numerous users who interact with \myvalue{paid} apps, quickly migrate to \myvalue{free} apps. 
Furthermore, it can be noticed from Figure~\ref{fig:migrations}(c) that users who interact with top-rated apps mostly do not compromise on app quality (i.e., the overall average rating of the app), which is depicted by very low migration to low-rated apps by those users. 
It is also clear that users, with a liking for highly rated apps, may migrate to apps with rating scores between \textit{3.5-4.5}. 
A very dynamic user migration pattern can be observed in Figure~\ref{fig:migrations}(d). 
Users migrate from one category to another among the top-5 categories quite dynamically. 
This highly fluctuating pattern depicts the complexity of modeling users' behavior in recommender systems, especially in cold-start settings such dynamic behavior becomes very challenging. 
It is important to note that {\ourdataset} has $48$ categories while in Figure~\ref{fig:migrations}(d), we show only top-5 categories with respect to the number of reviews. This dynamically fluctuating migration pattern can be even more complex if all the categories are considered.

\begin{figure}[t!]
    \centering
    \makebox[\linewidth][c]{\includegraphics[width=0.4\textwidth]{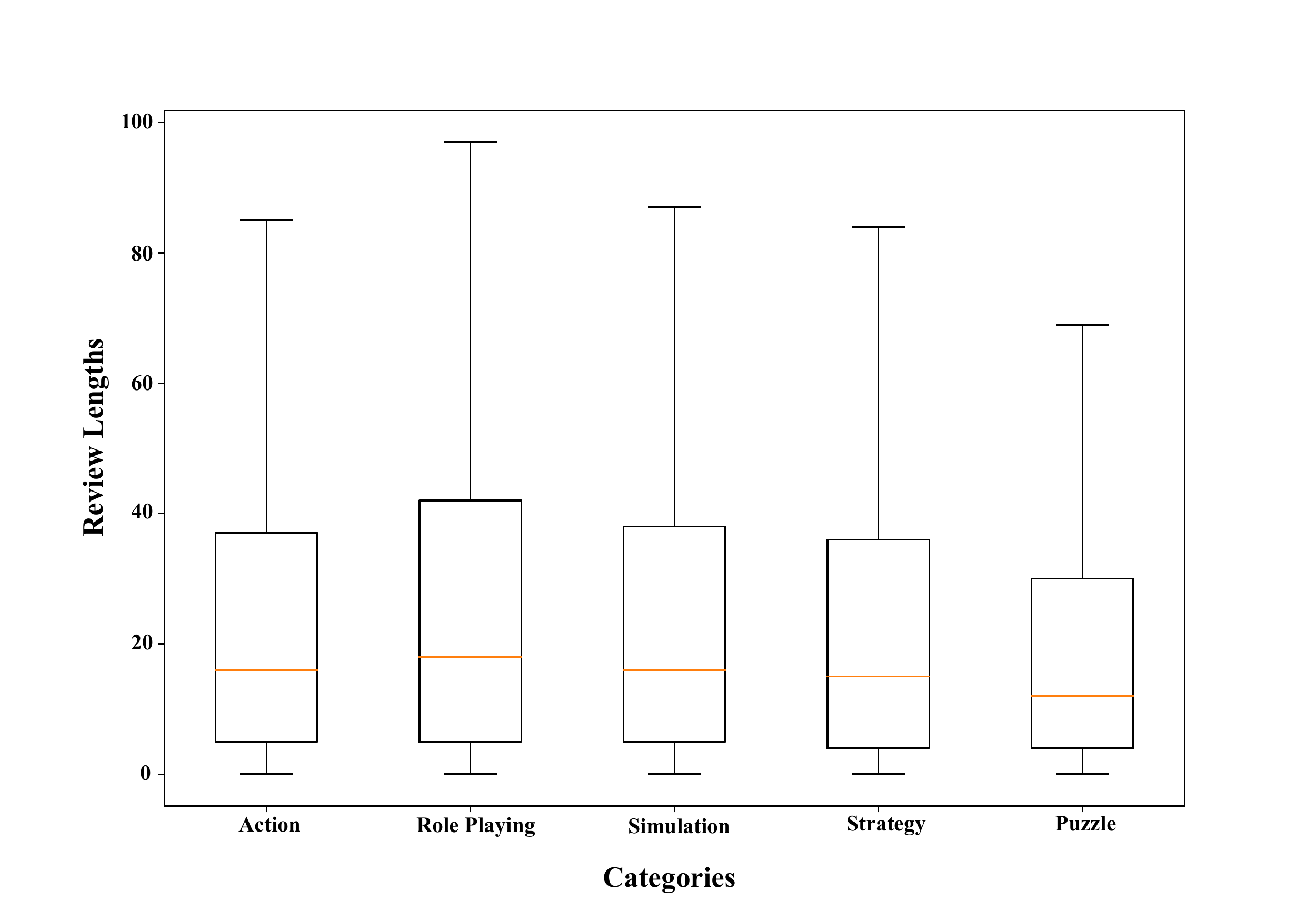}}
    \captionsetup{justification=centering}
    \caption{Review length for Top-5 categories based on the number of reviews.}
    \label{fig:top5_cat_review_len_boxplot}
    \vspace{-10pt}
\end{figure}

Figure \ref{fig:circular_bar_plot} summarizes the number of reviews in the top-10 categories. 
It can be observed that the \myvalue{Action} category has the most number of reviews among all the $48$ categories in the {\ourdataset} dataset. 
It is also interesting to note that the top-10 categories amount to roughly 50\% of the reviews in comparison with the rest of the categories.

Figure \ref{fig:top5_cat_review_len_boxplot} summarizes the review length in the top-5 categories: \myvalue{Action}, \myvalue{Role}~\myvalue{Playing}, \myvalue{Simulation}, \myvalue{Strategy}, and \myvalue{Puzzle}. 
We turn off the whiskers for better visibility in the box plot in Figure \ref{fig:top5_cat_review_len_boxplot}, but we provide some statistics about the outliers in each category.
For example, in the \myvalue{Action} category, there are $1,373,484$ total reviews with $10,357$ reviews having more than $100$ words. 
Hence, the outliers amount to nearly $0.754\%$ of the total reviews in the \myvalue{Action} category. 
Similarly, in the \myvalue{Role}~\myvalue{Playing} category, there are $1,321,861$ reviews in total with $12,187$ having more than 100 words giving rise to $0.922\%$ outliers. 
The \myvalue{simulation} category has $1,023,366$ reviews with $8,056$ having more than 100 words, which is $0.787\%$ of the total reviews in the category. 
The \myvalue{Strategy} category has $0.771\%$ reviews with more than $100$ words, which amounts to $7430$ reviews out of $963,360$ total reviews belonging to this category. 
Finally, the \myvalue{Puzzle} category has $954,909$ total reviews with $4,832$ reviews having more than $100$ words which are a marginal $0.506\%$ of the total reviews in the category. 
From Figure \ref{fig:top5_cat_review_len_boxplot}, it can be noticed that most of the reviews are less than $100$ words long in the top-5 categories. 

\subsection{Additional Usage Scenarios}
On top of building recommender systems, this large-scale dataset can be used for a variety of purposes. 
From app development and optimization to market research and fraud detection, the insights provided by this data can help businesses and researchers make informed decisions and create more effective strategies. 
In this work, we focus on providing baseline results for a wide range of recommender systems and leave other uses of this dataset to the research community.

As with any dataset, there are potential bad use cases for this dataset. 
As such, the data does not contain any information about the users.
Specifically, we de-identified and anonymized data to prevent individual identification.
For completeness, we consider it important to highlight potential bad uses that could arise.
For example, a company could pay for fake reviews or ratings to artificially boost their app's standing, or they could manipulate the app metadata to make their app be recommended more often by a certain recommender system.
Similarly, attackers could use the data to identify apps more favorable to be recommended and then create fake versions that include malicious code.

On balance, we believe that this dataset is more useful than the potential harm.
For example, this dataset has the potential to provide many useful insights and benefits for individuals and organizations. 
Specifically, we expect that this dataset will inspire research and development of app recommender systems, among other good uses.

\section{Baselines}
\label{sec:baseline}

We consider a wide range of recommender systems for benchmarking and establish baseline results for the {\ourdataset} dataset.

\subsection{General Recommendations}
\stitle{Pop.}
It is a simple popularity-based model. 
In this model, the popularity of items is recorded, and the most popular items are recommended to the user. 
For example, utilizing this model on Youtube, most watched videos will be recommended to users.

\subsection{Sequential Recommendations}
\stitle{SASRec~\cite{kang2018self}.} Markov chains assume that a user's next action can be determined by their recent actions (just one or few recent actions). On the other hand, RNN-based methods can consider long-term user-item interaction history for uncovering the hidden interests of a user. SASRec combines the best of Markov Chain (MC) based methods and Recurrent Neural Network (RNN) based methods in a unified design by employing the attention mechanism. MC-based approaches can be effective where the sparse dataset is involved, while RNN-based approaches are good for situations involving denser datasets. SASRec proposes to balance the power of MC-based approaches and RNN-based approaches to get the best of both worlds. 

\stitle{ELECRec~\cite{chen2022elecrec}.} Next item prediction task is generally modeled as a generative task. {ELECRec} proposes to employ a discriminator in the recommender system for the next item prediction task. The discriminator is responsible for deciding if the next sampled item is a true target item. A generator $G$ and discriminator $D$ are trained jointly. The generator $G$ is trained to generate high-quality samples for the discriminator.

\stitle{BERT4Rec~\cite{sun2019bert4rec}.}
Building upon the observation that sequential neural networks are sub-optimal and limited. This sub-optimality can be attributed to restrictive unidirectional (left to right) encoding of the user's behavior and their assumption is that the user-interaction sequence is ordered. BERT4Rec is a sequential recommendation model which employs the bidirectional self-attention for encoding the user's interaction sequence. This baseline adopts the cloze objective to the sequential recommendation and predicts the masked items by joint conditioning on their bidirectional context.

\stitle{HGN~\cite{ma2019hierarchical}.}
Considering the importance of recent chronological user-item interaction, the hierarchical gating network proposes a method to capture both long-term and short-term user interests by integrating Bayesian Personalized Ranking (BPR). The proposed method consists of a feature gating module, item gating module, and item-item product module. The gating module is responsible to decide which features are to be forwarded to downstream layers from the feature and instance layers. 
The item-item product module captures the item relations between items that the user interacted with in the past and will interact with in the future.

\stitle{SINE~\cite{tan2021sparse}.}
Realizing the importance of multiple conceptually distinct items in a user's behavior sequence, SINE suggests that capturing a unified embedding for a user's behavior sequence is affected primarily by the most recent user interactions. Building upon this observation, SINE proposes to use multiple embeddings to capture various aspects of a user's interaction behavior. Since the concept pool can be large, SINE has the ability to infer a sparse set of concepts from the large concept pool. Having multiple embeddings, an interest aggregation module is employed for predicting a user's current intention, which is also used for modeling next-item prediction. 

\subsection{Transformer Based Recommendation}
\stitle{LightSANs~\cite{fan2021lighter}.}
Self-attention networks (SANs) are limited due to quadratic complexity, and vulnerability to over-parameterization. Modeling inaccuracy in sequential relations between items is also a limiting factor in SANs because of implicit position encoding. To this end, LightSAN proposes a low-rank decomposed self-attention network. Low-rank decomposed self-attention projects users' historical items into a few latent  interests. Item-to-interest interaction is used to generate the context-aware representation.

\subsection{Session Based Recommendation}
\stitle{GRU4Rec~\cite{tan2016improved}.}
This baseline uses an RNN-based method for the session-based recommendation. Data augmentation and method to account for the distribution shift in the input data is employed in the proposed technique. 

\subsection{GNN Based Recommendation}
\stitle{GCSAN~\cite{xu2019graph}.}
This method proposes a graph-contextualized self-attention model which employs both graph neural networks and self-attention networks. In GCSAN, rich local dependencies are captured with graph neural networks, while long-range dependencies are captured with the self-attention network. Each session is represented as a combination of global preference and the current interest of the session.
\section{Experimental Setup}
\label{sec:experiments}

\subsection{Experimental Settings}
For sequential recommendation baselines, we keep a consistent batch size of $4096$ and the maximum interaction length to be $50$. 
We employ a leave-one-out strategy for validation and testing, and the full item set is used for evaluation. We use early stopping patience of $10$. We consider the sequential recommendation task as a multiclass classification task and use the cross-entropy loss for training the models. {SASRec~\cite{kang2018self}} is trained with {adam} optimizer and a learning rate of 0.001. The number of layers and attention heads is $2$. The dropout rate is $0.5$ and {gelu} is used as an activation function. {HGN~\cite{ma2019hierarchical}} and {SINE~\cite{tan2021sparse}} are trained with embedding size $64$, learning rate $0.001$, and {adam} optimizer. {LigthSANs~\cite{fan2021lighter}} uses the latent interest dimension to be $5$, $2$ attention heads and $2$ transformer layers. Training is done with a learning rate of $0.001$ and {adam} optimizer. GCSAN~\cite{xu2019graph} has $2$ transformer encoder layers, $2$ attention heads, 64 is the hidden state features size, feed-forward layers' hidden size is $256$, weight is set to be $0.6$ and the number of layers in graph neural network is $1$, {adam} is used for optimization with a learning rate of $0.001$. GRU4Rec~\cite{tan2016improved} is trained with $1$ layer with an embedding size of $64$, hidden size of $128$, and dropout to be $0.3$. We employ RecBole~\cite{recbole[1.0]} for establishing the baselines for \ourdataset.

\subsection{Evaluation Metrics}
We employ standard metrics, {Hit@K} and {NDCG@K} where $K \in \{1,5,10,15,20\}$, as our evaluation metrics for the benchmark methods. 
{Hit@k} considers the number of times the predicted item appears in the top $K$ list and can be represented as in ~\cite{li2020sampling}:
$$
    HR@K=\frac{1}{M}\sum_{u=1}^{M}\mathbbm{1}_{R_u\leq K}  = \sum_{R=1}^{N}W_R \cdot \mathbbm{1}_{R\leq K}
$$

$$
    W_R = \frac{1}{M}\sum_{u=1}^{M}\mathbbm{1}_{R_u=R}
$$
$\mathbbm{1}_X$ represents the indicator random variable, $M$ are the total number of users and $N=|I|$ are the total items. R is the integer rank position of an item in the range $[1, N]$. $R_u$ is the rank of item $i_u$ among $I$ items, for user $u$. $W_R$ captures the users with item $i_u$ at position $R$.

\textit{NDCG@K} can be represented as ~\cite{jarvelin2002cumulated, recbole[1.0]}:
$$
    \frac{1}{|U|} \sum_{u \in U}\left(\frac{1}{\sum_{i=1}^{\min (|R(u)|, K)} \frac{1}{\log _2(i+1)}} \sum_{i=1}^K \delta(i \in R(u)) \frac{1}{\log _2(i+1)}\right)
$$

All the item set is considered for ranking the prediction.
\section{Results and Discussion}
\label{sec:results}

\begin{table*}[h]
\centering
\caption{Performance Analysis of various baselines on {\ourdataset}. 
\label{tbl:perf-analysis}
Baselines belonging to various categories, such as General baselines (e.g. Pop), Sequential baselines (e.g. SASRec, ELECRec, BERT4Rec, HGN, SINE), Session-based baselines (e.g. GRU4Rec GCSAN), Graph Neural Network based baselines (e.g. GCSAN), Transformer based baselines (e.g. LigthSANs).}

\begin{tabular}{c|ccccccccc}
\toprule
\multicolumn{1}{l}{\textbf{Method $\downarrow$ Metric $\rightarrow$}} & \multicolumn{1}{l}{\textbf{Hit@1}} & \multicolumn{1}{l}{\textbf{Hit@5}} & \multicolumn{1}{l}{\textbf{Hit@10}} & \multicolumn{1}{l}{\textbf{Hit@15}} & \multicolumn{1}{l}{\textbf{Hit@20}} & \multicolumn{1}{l}{\textbf{NDCG@5}} & \multicolumn{1}{l}{\textbf{NDCG@10}} & \multicolumn{1}{l}{\textbf{NDCG@15}} & \multicolumn{1}{l}{\textbf{NDCG@20}} \\ \hline
\textbf{Pop}                                & 0.0027                             & 0.0086                             & 0.0151                              & 0.0208                              & 0.0256                              & 0.0056                              & 0.0077                               & 0.0092                               & 0.0103                               \\
\textbf{SASRec}                             & 0.0026                             & 0.0098                             & 0.0181                              & 0.0242                              & 0.0295                              & 0.0061                              & 0.0088                               & 0.0104                               & 0.0117                               \\
\textbf{ElecRec}                            & 0.0020                             & 0.0094                             & 0.0174                              & 0.0237                              & 0.0293                              & 0.0056                              & 0.0082                               & 0.0098                               & 0.0112                               \\
\textbf{Bert4Rec}                           & 0.0024                             & 0.0083                             & 0.014                               & 0.0183                              & 0.0221                              & 0.0054                              & 0.0072                               & 0.0083                               & 0.0092                               \\
\textbf{HGN}                                & 0.0012                             & 0.0054                             & 0.0096                              & 0.0132                              & 0.0165                              & 0.0033                              & 0.0046                               & 0.0056                               & 0.0064                               \\
\textbf{SINE}                               & 0.0022                             & 0.0087                             & 0.0163                              & 0.0228                              & 0.028                               & 0.0054                              & 0.0078                               & 0.0095                               & 0.0107                               \\
\textbf{LightSANs}                          & 0.0024                             & 0.0102                             & 0.0172                              & 0.0227                              & 0.028                               & 0.0062                              & 0.0085                               & 0.0099                               & 0.0112                               \\
\textbf{GRU4Rec}                            & 0.0021                             & 0.0086                             & 0.0153                              & 0.021                               & 0.0261                              & 0.0053                              & 0.0074                               & 0.0089                               & 0.0102                               \\
\textbf{GCSAN}                              & 0.0024                             & 0.0094                             & 0.0161                              & 0.0214                              & 0.0266                              & 0.0059                              & 0.0081                               & 0.0095                               & 0.0107  \\
\bottomrule
\end{tabular}
\end{table*}
Table~\ref{tbl:perf-analysis} presents the performance analysis of various baselines on \ourdataset. Next, we discuss these results and provide further details.
%
As discussed earlier, Pop is a popularity-based model which relies on the popularity of items. Pop is the most naive model and, as expected, its performance is the worst compared to almost all of the other baselines. 
SASRec had previously reported better results than Pop according to Hit@10 and NDCG@10 performance metrics. 
In SASRec, 100 negative items were randomly sampled, and {Hit@10} and {NDCG@10} was calculated against these 101 items, including the ground truth. 
Using {\ourdataset}, we employ a full item set for evaluation, which is a stricter evaluation criterion for recommendation systems. 
We observe that on all the performance metrics, SASRec outperforms Pop except {Hit@1}. 
For example, SASRec manages to get $19.86\%$ improved {Hit@10} in comparison with pop. 
On the same lines, a $16.34\%$ performance improvement is observed in {Hit@15}. 
Similarly, SASRec achieves $15.23\%$ improved Hit@20 in comparison with Pop on \ourdataset. 

Considering {NDCG} metric, SASRec had reported $41.37\%$ improved {NDCG@10} compared to pop, when evaluation is done on \myvalue{Beauty}. 
On \ourdataset, we observe only $14.28\%$ improvement in {NDCG@10} compared to pop. 
We believe that this difference in the improvement on {NDCG@10} between SASRec and Pop on these two datasets (\myvalue{Beauty} and \ourdataset) can be explained by keeping in view the comparative dataset statistics of the \myvalue{Beauty} and \ourdataset. 
There are $52,024$ users in \myvalue{Beauty} while \ourdataset has $700,111$ users. Secondly, \myvalue{Beauty} dataset has $57,289$ items compared to {\ourdataset} which has $10,173$ apps. \myvalue{Beauty} has $0.4M$ interactions, with $7.6$ average interaction per user and $6.9$ average interaction per item. 
Whereas {\ourdataset} has $19.3M$ interactions with $27.56$ interactions per user and $1,896.88$ interactions per item (i.e., app). 
For example, {\ourdataset} has $27.56$ interaction per user, which is more than 200 times more interactions per user than \myvalue{Beauty}. Likewise, {\ourdataset} also has a considerably larger number of average interactions per item in comparison with \myvalue{Beauty}. Keeping these factors in view, {\ourdataset} presents a more dynamic recommendation scenario. 
The high degree of user and item interactions in {\ourdataset} can be attributed to being the reason for the comparatively smaller improvement in {NDCG@10} by  SASRec compared to the improvement SASRec had reported over Pop on \myvalue{Beauty}. Secondly, since we consider a full item set for evaluation compared to the ranking strategy by SASRec, this may also be a potential reason for smaller performance gains by SASRec in comparison with pop. 
On \myvalue{Beauty}, SASRec reported better results than GRU4Rec. We observe the same pattern of SASRec outclassing the GRU4Rec when training and evaluation are performed on {\ourdataset}.

On \myvalue{Beauty} dataset, ELECRec reports better {Hit@5}, {Hit@10}, {NDCG@5} and {NDCG@10} than pop. We notice the same pattern when training and evaluation are done on {\ourdataset} with {Hit@1} to be the only exception, where Pop performs better than ELECRec. 
We assume that a very dynamic user-item interaction history might be the reason. 
Similarly, ELECRec had reported $329.87\%$ improved {Hit@5} and  $242.04\%$ improvement in {Hit@10} compared to GRU4Rec on \myvalue{Beauty}. 
Moreover, for {NDCG@5} and {NDCG@10} metrics, ELECRec achieved $412.12\%$ and $331.38\%$ improvements over GRU4Rec, respectively. 
Focusing on the performance improvements obtained by ELECRec over GRU4Rec when training and evaluation are done on {\ourdataset}, ELECRec manages to get a $9.30\%$ and $13.72\%$ improvement in {Hit@5} and {Hit@10}, respectively. 
Similarly, $1.88\%$ and  $10.81\%$ improvement is observed in {NDCG@5} and {NDCG@10}, respectively. 
This observation also points toward the highly dynamic user-item interaction involved in {\ourdataset}. 
Both models struggle to achieve high performance, which results in diminished performance gaps between the two approaches as compared to the performance gaps reported in ELECRec. 
There are over 19 million interactions in {\ourdataset} while only 0.4M user-item interactions are there in \myvalue{Beauty}. 
This high level of dynamism in user-item interactions pushes the models towards lower performance gains. 
Considering this analysis, {\ourdataset} should be a valuable addition to the existing volume of recommendation datasets. 

ELECRec also shows the comparative gains over SASRec on \myvalue{Beauty}. First, we will cover the relative performance gains ELECRec manages to get over SASRec with respect to {Hit@5}, {Hit@10}, {NDCG@5} and {NDCG@10} on \myvalue{Beauty} dataset. After that, we will discuss the performance improvements that ELECRec achieves over SASRec when the training and evaluation dataset is {\ourdataset}. Starting off with {Hit@5} and {Hit@10}, ELECRec reports $83.59\%$ and $59.47\%$ improvement over SASRec. Similarly, $103.61\%$ and $84.11\%$ improvement is reported by ELECRec over SASRec with respect to {NDCG@5} and {NDCG@10}, respectively. 
Now, let us discuss the performance comparison between ELECRec and SASRec when training and evaluation are performed on {\ourdataset}. 
It can be observed that SASRec performs better than ELECRec on {\ourdataset} in {Hit@5}, {Hit@10}, {NDCG@5}, and {NDCG@10} metrics. 
The quantification of the performance improvement of SASRec over ELECRec in {Hit@5}, {Hit@10}, {NDCG@5}, and {NDCG@10} yields $4.25\%$, $4.02\%$, $8.92\%$, and $7.31\%$, respectively. 
We believe that ELECRec employs a generator that is trained with an NLP task, while the discriminator is responsible for discerning if the item is the next rightful (real) item in the sequence or a fake next item. The discriminators discerning ability depends upon the quality of samples generated by the generator. Since {\ourdataset} presents as highly dynamic user-item interaction sequences with fleeting interests shown by users, the generator might have difficulty generating high-quality training samples. Since the discriminators' ability to capture the true item correlation depends upon the quality of samples generated by the generator, low-quality samples may lead to diminished performance gains on the discriminator's part. 

It is worth recalling Figure~\ref{fig:migrations} that provides an interesting snapshot of the dynamic user-item interaction sequences. 
For example, it can be noticed from Figure~\ref{fig:migrations}(b) that a considerable percentage of users migrate from paid applications to free applications. 
A similar dynamic migration pattern is evident from Figure~\ref{fig:migrations}(d) which depicts the user migration among top-5 categories. 
We think that with 19 million user-item interactions having such a dynamic temporal interaction pattern may present a challenge for the generator to learn high-quality sample generation. Moreover, ELECRec reported its comparative performance to BERT4Rec. 
ELECRec outperformed BERT4Rec in the original paper on all the metrics on \myvalue{Beauty}~\cite{chen2022elecrec}. 
Specifically, ELECRec had reported a $100.85\%$ and $61.06\%$ performance gain over BERT4Rec in {Hit@5} and {Hit@10}, respectively, and $131.50\%$ and $97\%$ gains were reported in {NDCG@5} and {NDCG@10} metrics. 
We observe the same pattern of ELECRec outperforming BERT4Rec in most of the evaluation metrics when {\ourdataset} is employed for training and evaluation except for {Hit@1} where BERT4Rec does better than ELECRec. 
Considering the fact that {Hit@1} is a stricter criterion, we trust that the training objective of  predicting the masked items in the sequence by joint conditioning on bidirectional context, employed by BERT4Rec might be a better training objective for stricter evaluation metrics like {Hit@1}. 
Especially, in the context of fleeting user interests like in {\ourdataset}, masked item prediction objective  with joint conditioning on bidirectional context seems to be effective for encoding the user's dynamic behavior.

Graph contextualized self-attention model GCSAN employs both graph neural networks and self-attention networks for the session-based recommendation. 
In the original paper, several comparative evaluations are reported between GCSAN and competing methods. We are primarily interested in the quantification of the comparative representation learning ability of GRU4Rec and Pop versus GCSAN since we also include Pop and  GRU4Rec in our baselines along with GCSAN. 
In the original paper, using {Amazon-Books} dataset,  GCSAN reports improvement over Pop and GRU4Rec in {NDCG@5} and {NDCG@10}. 
Similarly, improved results are reported as compared to Pop and GRU4Rec in {Hit@5} and {Hit@10} on different benchmark datasets. We observe similar pattern of GCSAN outperforming Pop and GRU4Rec in {NDCG@5}, {NDCG@10}, {Hit@5}, and {Hit@10} consistently on {\ourdataset}. We notice $9.30\%$ and $6.62\%$ improvement over Pop in {NDCG@5} and {NDCG@10}, respectively. 
Similarly, improvement is observed considering {Hit@5}, {Hit@10}, {NDCG@5}, and {NDCG@10} for  GCSAN over GRU4Rec.

LightSANs is a transformer variant for the next-item recommendation task which employs a low-rank decomposed self-attention for projecting the user's historical interests into a few latent interests. LightSANs reported {Hit@10} and {NDCG@10} results on several competing baselines. 
We will focus on Pop,  GRU4Rec, BERT4Rec, and SASRec for comparing and analyzing the LightSANs performance on {Amazon-Books} dataset and {\ourdataset}. 
Before further going into the comparative analysis of the performance of LightSANs on {Amazon-Beauty} versus {\ourdataset}, let us first look at the {Amazon-Books} datasets. 
{Amazon-Books} dataset has $19$K users, $60$K items, and $1.7$ million interactions. 
On {Amazon-Books} dataset, LightSANs reports a $121.77\%$ and $172.43\%$ improvement in {Hit@10} and {NDCG@10} over Pop. 
Similarly, $3.91\%$ and $2.65\%$ of improvement is reported in {Hit@10} and {NDCG@10}, respectively, over SASRec. 
$8.41\%$ and $5.72\%$ improved {Hit@10} and {NDCG@10} are reported compared to those of GRU4Rec. 
Finally, an improvement of $8.76\%$ and $4.03\%$ is reported over BERT4Rec. 
When dataset is {\ourdataset}, we observe $13.90\%$ and $10.38\%$ improvement over Pop in {Hit@10} and {NDCG@10}. LightSANs manages to get $12.41\%$ and $14.86\%$ improvement over GRU4Rec for the same metrics.

Similar to ELECRec, LightSANs also shows $22.85\%$ and $18.05\%$ better results than BERT4Rec for {Hit@10} and {NDCG@10} metrics. 
We notice that LightSANs maintains the pattern of outperforming competing baselines on {\ourdataset} as it does on the Amazon-Books dataset, with SASRec being an exception that outperforms LightSANs in {Hit@10} and {NDCG@10} metrics. 
The consistent performance exhibited by LightSANs against BERT4Rec, Pop, and GRU4Rec can be attributed to better design choices in LightSANs. For example, LightSANs proposes a decoupled position encoding in place of implicit position encoding. We believe that this implicit position encoding helps LightSANs to model the historical user interests and user-item interaction more precisely, which steers the model towards better {Hit@10} and {NDCG@10} on {\ourdataset}. Nevertheless, LightSANs is outperformed by SASRec. The superior performance of SASRec might be due to the learnable position embeddings introduced by SASRec. 

SINE investigates the idea of encoding a user's interests with multiple embedding vectors, building upon their empirical findings that a user's behavior sequence exhibits multiple distinct interests. SINE employs several datasets for benchmarking their results. We consider the performance reported by SINE on the Amazon Product Review dataset using {Hit@10} and {NDCG@10} metrics. We restrict our comparative analysis to SASRec and GRU4Rec. SINE reports improvement over SASRec in Hit@50, Hit@100, NDCG@50, and NDCG@100, respectively. Better performance is also reported against GRU4Rec in Hit@50, Hit@100, NDCG@50, and NDCG@100. 
Given {\ourdataset} for training and evaluation, SINE manages to outclass GRU4Rec by $7.27\%$ and $4.90\%$ in Hit@20 and NDCG@20, respectively; but struggles to perform better than SASRec on Hit@20 and NDCG@20. 
First, it might be because Hit@20 and NDCG@20 are stricter criteria as compared to Hit@50, Hit@100, NDCG@50, and NDCG@100 opted by SINE. 
Secondly, keeping in view that SINE strives to capture the user's distinct interests from the interaction history of the user, we think that converging to distinct user interests in {\ourdataset} presents a challenge because of fluctuating user interests, as depicted in Figure \ref{fig:migrations}. 
This inability of learning and embedding the distinct user interests from the interaction sequence may lead SINE to show worse performance than SASRec. 

\section{Conclusion}
\label{sec:conclusion}
In this paper, we have introduced {\ourdataset}, a large-scale dataset of sequential user-app interactions.
The unique feature of {\ourdataset} is that it captures multiple interactions per user, providing a more comprehensive view of user behavior. 
With a total of $19.3$ million user-app interactions, spanning across more than 10 thousand distinct apps from 48 categories, and involving $0.7$ million unique users, each with at least five distinct app interactions, {\ourdataset} offers unprecedented granularity in understanding user engagement across different app categories. 
Moreover, every user-app interaction in {\ourdataset} contains rich contextual information, such as the user's rating, textual review, and review date. 
Last but not least, each app in {\ourdataset} carries extensive metadata, such as the app's title, category, long-form textual description, overall average rating, developer information, and content rating, among others.
We also show the usefulness of the {\ourdataset} dataset as an experimental testbed for research in app recommendation by conducting a comparative evaluation of various state-of-the-art recommendation techniques. 
This evaluation also establishes baseline results that will benefit the research community.
We hope that our dataset will inspire further research, enable novel insights, and pave the way for future mobile app recommendation systems.
\balance
\bibliographystyle{plain}
\bibliography{sample-base.bib}

\begin{thebibliography}{10}

\bibitem{movielens}
Movielens.
\newblock \url{https://grouplens.org/datasets/movielens/}, 2022.
\newblock Accessed: 2022-11-06.

\bibitem{selenium}
Selenium webdrive.
\newblock \url{https://www.selenium.dev/documentation/webdriver/}, 2022.
\newblock Accessed: 2023-18-02.

\bibitem{top20-dataset}
Top 20 play store app reviews.
\newblock
  \url{https://www.kaggle.com/datasets/odins0n/top-20-play-store-app-reviews-daily-update},
  2022.
\newblock Accessed: 2022-12-09.

\bibitem{yelp-dataset}
Yelp open dataset.
\newblock \url{https://www.yelp.com/dataset}, 2022.
\newblock Accessed: 2023-18-02.

\bibitem{appstore}
Apple.
\newblock Apple app store.
\newblock \url{https://apps.apple.com/}, 2022.
\newblock Accessed: 2022-11-06.

\bibitem{number-of-apps-on-stores}
L.~Ceci.
\newblock Number of apps available in leading app store.
\newblock
  \url{https://www.statista.com/statistics/276623/number-of-apps-available-in-leading-app-stores/}.
\newblock Accessed: 2022-11-06.

\bibitem{chen2022elecrec}
Yongjun Chen, Jia Li, and Caiming Xiong.
\newblock Elecrec: Training sequential recommenders as discriminators.
\newblock {\em arXiv preprint arXiv:2204.02011}, 2022.

\bibitem{number-of-smartphone-users}
J.~Degenhard.
\newblock Number of apps available in leading app store.
\newblock
  \url{https://www.statista.com/forecasts/1143723/smartphone-users-in-the-world}.
\newblock Accessed: 2022-02-02.

\bibitem{devlin2018bert}
Jacob Devlin, Ming-Wei Chang, Kenton Lee, and Kristina Toutanova.
\newblock Bert: Pre-training of deep bidirectional transformers for language
  understanding.
\newblock {\em arXiv preprint arXiv:1810.04805}, 2018.

\bibitem{diao2014jointly}
Qiming Diao, Minghui Qiu, Chao-Yuan Wu, Alexander~J Smola, Jing Jiang, and
  Chong Wang.
\newblock Jointly modeling aspects, ratings and sentiments for movie
  recommendation (jmars).
\newblock In {\em Proceedings of the 20th ACM SIGKDD international conference
  on Knowledge discovery and data mining}, pages 193--202, 2014.

\bibitem{elman1990finding}
Jeffrey~L Elman.
\newblock Finding structure in time.
\newblock {\em Cognitive science}, 14(2):179--211, 1990.

\bibitem{fan2021lighter}
Xinyan Fan, Zheng Liu, Jianxun Lian, Wayne~Xin Zhao, Xing Xie, and Ji-Rong Wen.
\newblock Lighter and better: low-rank decomposed self-attention networks for
  next-item recommendation.
\newblock In {\em Proceedings of the 44th international ACM SIGIR conference on
  research and development in information retrieval}, pages 1733--1737, 2021.

\bibitem{aarsynth}
Umar Farooq, AB~Siddique, Fuad Jamour, Zhijia Zhao, and Vagelis Hristidis.
\newblock App-aware response synthesis for user reviews.
\newblock In {\em 2020 IEEE International Conference on Big Data (Big Data)},
  pages 699--708. IEEE, 2020.

\bibitem{pprior}
Moghis Fereidouni, Adib Mosharrof, Umar Farooq, and AB~Siddique.
\newblock Proactive prioritization of app issues via contrastive learning.
\newblock In {\em 2022 IEEE International Conference on Big Data (Big Data)},
  pages 535--544. IEEE, 2022.

\bibitem{rrgen}
Cuiyun Gao, Jichuan Zeng, Xin Xia, David Lo, Michael~R Lyu, and Irwin King.
\newblock Automating app review response generation.
\newblock In {\em 2019 34th IEEE/ACM International Conference on Automated
  Software Engineering (ASE)}, pages 163--175. IEEE, 2019.

\bibitem{googleplay}
Google.
\newblock Google play store.
\newblock \url{https://play.google.com/store/apps}, 2022.
\newblock Accessed: 2022-11-06.

\bibitem{harper2015movielens}
F~Maxwell Harper and Joseph~A Konstan.
\newblock The movielens datasets: History and context.
\newblock {\em Acm transactions on interactive intelligent systems (tiis)},
  5(4):1--19, 2015.

\bibitem{he2016ups}
Ruining He and Julian McAuley.
\newblock Ups and downs: Modeling the visual evolution of fashion trends with
  one-class collaborative filtering.
\newblock In {\em proceedings of the 25th international conference on world
  wide web}, pages 507--517, 2016.

\bibitem{hochreiter1997long}
Sepp Hochreiter and J{\"u}rgen Schmidhuber.
\newblock Long short-term memory.
\newblock {\em Neural computation}, 9(8):1735--1780, 1997.

\bibitem{iacob2013retrieving}
Claudia Iacob and Rachel Harrison.
\newblock Retrieving and analyzing mobile apps feature requests from online
  reviews.
\newblock In {\em 2013 10th working conference on mining software repositories
  (MSR)}, pages 41--44. IEEE, 2013.

\bibitem{iwata2011fashion}
Tomoharu Iwata, Shinji Watanabe, and Hiroshi Sawada.
\newblock Fashion coordinates recommender system using photographs from fashion
  magazines.
\newblock In {\em Twenty-Second International Joint Conference on Artificial
  Intelligence}, 2011.

\bibitem{jarvelin2002cumulated}
Kalervo J{\"a}rvelin and Jaana Kek{\"a}l{\"a}inen.
\newblock Cumulated gain-based evaluation of ir techniques.
\newblock {\em ACM Transactions on Information Systems (TOIS)}, 20(4):422--446,
  2002.

\bibitem{kang2018self}
Wang-Cheng Kang and Julian McAuley.
\newblock Self-attentive sequential recommendation.
\newblock In {\em 2018 IEEE international conference on data mining (ICDM)},
  pages 197--206. IEEE, 2018.

\bibitem{khalid2013identifying}
Hammad Khalid.
\newblock On identifying user complaints of ios apps.
\newblock In {\em 2013 35th international conference on software engineering
  (ICSE)}, pages 1474--1476. IEEE, 2013.

\bibitem{khalid2014mobile}
Hammad Khalid, Emad Shihab, Meiyappan Nagappan, and Ahmed~E Hassan.
\newblock What do mobile app users complain about?
\newblock {\em IEEE software}, 32(3):70--77, 2014.

\bibitem{koren2008factorization}
Yehuda Koren.
\newblock Factorization meets the neighborhood: a multifaceted collaborative
  filtering model.
\newblock In {\em Proceedings of the 14th ACM SIGKDD international conference
  on Knowledge discovery and data mining}, pages 426--434, 2008.

\bibitem{li2020sampling}
Dong Li, Ruoming Jin, Jing Gao, and Zhi Liu.
\newblock On sampling top-k recommendation evaluation.
\newblock In {\em Proceedings of the 26th ACM SIGKDD International Conference
  on Knowledge Discovery \& Data Mining}, pages 2114--2124, 2020.

\bibitem{ma2019hierarchical}
Chen Ma, Peng Kang, and Xue Liu.
\newblock Hierarchical gating networks for sequential recommendation.
\newblock In {\em Proceedings of the 25th ACM SIGKDD international conference
  on knowledge discovery \& data mining}, pages 825--833, 2019.

\bibitem{maalej2015bug}
Walid Maalej and Hadeer Nabil.
\newblock Bug report, feature request, or simply praise? on automatically
  classifying app reviews.
\newblock In {\em 2015 IEEE 23rd international requirements engineering
  conference (RE)}, pages 116--125. IEEE, 2015.

\bibitem{amazon-dataset}
Julian McAuley.
\newblock Amazon product data.
\newblock \url{http://jmcauley.ucsd.edu/data/amazon/}, 2022.
\newblock Accessed: 2022-11-06.

\bibitem{mcauley2015image}
Julian McAuley, Christopher Targett, Qinfeng Shi, and Anton Van Den~Hengel.
\newblock Image-based recommendations on styles and substitutes.
\newblock In {\em Proceedings of the 38th international ACM SIGIR conference on
  research and development in information retrieval}, pages 43--52, 2015.

\bibitem{mcilroy2016analyzing}
Stuart McIlroy, Nasir Ali, Hammad Khalid, and Ahmed E.~Hassan.
\newblock Analyzing and automatically labelling the types of user issues that
  are raised in mobile app reviews.
\newblock {\em Empirical Software Engineering}, 21:1067--1106, 2016.

\bibitem{ni2019justifying}
Jianmo Ni, Jiacheng Li, and Julian McAuley.
\newblock Justifying recommendations using distantly-labeled reviews and
  fine-grained aspects.
\newblock In {\em Proceedings of the 2019 conference on empirical methods in
  natural language processing and the 9th international joint conference on
  natural language processing (EMNLP-IJCNLP)}, pages 188--197, 2019.

\bibitem{ni19modeling}
Jianmo Ni, Larry Muhlstein, and Julian McAuley.
\newblock Modeling heart rate and activity data for personalized fitness
  recommendation.
\newblock In {\em WWW}, 2019.

\bibitem{radford2019language}
Alec Radford, Jeff Wu, Rewon Child, David Luan, Dario Amodei, and Ilya
  Sutskever.
\newblock Language models are unsupervised multitask learners.
\newblock 2019.

\bibitem{rappaz2021recommendation}
J{\'e}r{\'e}mie Rappaz, Julian McAuley, and Karl Aberer.
\newblock Recommendation on live-streaming platforms: Dynamic availability and
  repeat consumption.
\newblock In {\em Proceedings of the 15th ACM Conference on Recommender
  Systems}, pages 390--399, 2021.

\bibitem{rappaz21recommendation}
Jérémie Rappaz, Julian McAuley, and Karl Aberer.
\newblock Recommendation on live-streaming platforms: Dynamic availability and
  repeat consumption.
\newblock In {\em RecSys}, 2021.

\bibitem{srisopha-how}
Kamonphop Srisopha, Daniel Link, and Barry Boehm.
\newblock How should developers respond to app reviews? features predicting the
  success of developer responses.
\newblock EASE 2021, page 119–128, New York, NY, USA, 2021. Association for
  Computing Machinery.

\bibitem{sun2019bert4rec}
Fei Sun, Jun Liu, Jian Wu, Changhua Pei, Xiao Lin, Wenwu Ou, and Peng Jiang.
\newblock Bert4rec: Sequential recommendation with bidirectional encoder
  representations from transformer.
\newblock In {\em Proceedings of the 28th ACM international conference on
  information and knowledge management}, pages 1441--1450, 2019.

\bibitem{tan2021sparse}
Qiaoyu Tan, Jianwei Zhang, Jiangchao Yao, Ninghao Liu, Jingren Zhou, Hongxia
  Yang, and Xia Hu.
\newblock Sparse-interest network for sequential recommendation.
\newblock In {\em Proceedings of the 14th ACM International Conference on Web
  Search and Data Mining}, pages 598--606, 2021.

\bibitem{tan2016improved}
Yong~Kiam Tan, Xinxing Xu, and Yong Liu.
\newblock Improved recurrent neural networks for session-based recommendations.
\newblock In {\em Proceedings of the 1st workshop on deep learning for
  recommender systems}, pages 17--22, 2016.

\bibitem{tanjim20attentive}
Mehrab Tanjim, Congzhe Su, Ethan Benjamin, Diane Hu, Liangjie Hong, and Julian
  McAuley.
\newblock Attentive sequential models of latent intent for next item
  recommendation.
\newblock In {\em WWW}, 2020.

\bibitem{vaswani2017attention}
Ashish Vaswani, Noam Shazeer, Niki Parmar, Jakob Uszkoreit, Llion Jones,
  Aidan~N Gomez, {\L}ukasz Kaiser, and Illia Polosukhin.
\newblock Attention is all you need.
\newblock {\em Advances in neural information processing systems}, 30, 2017.

\bibitem{wan2018item}
Mengting Wan and Julian McAuley.
\newblock Item recommendation on monotonic behavior chains.
\newblock In {\em Proceedings of the 12th ACM conference on recommender
  systems}, pages 86--94, 2018.

\bibitem{wan2019fine}
Mengting Wan, Rishabh Misra, Ndapa Nakashole, and Julian McAuley.
\newblock Fine-grained spoiler detection from large-scale review corpora.
\newblock {\em arXiv preprint arXiv:1905.13416}, 2019.

\bibitem{wu2020mind}
Fangzhao Wu, Ying Qiao, Jiun-Hung Chen, Chuhan Wu, Tao Qi, Jianxun Lian,
  Danyang Liu, Xing Xie, Jianfeng Gao, Winnie Wu, et~al.
\newblock Mind: A large-scale dataset for news recommendation.
\newblock In {\em Proceedings of the 58th Annual Meeting of the Association for
  Computational Linguistics}, pages 3597--3606, 2020.

\bibitem{xu2019graph}
Chengfeng Xu, Pengpeng Zhao, Yanchi Liu, Victor~S Sheng, Jiajie Xu, Fuzhen
  Zhuang, Junhua Fang, and Xiaofang Zhou.
\newblock Graph contextualized self-attention network for session-based
  recommendation.
\newblock In {\em IJCAI}, volume~19, pages 3940--3946, 2019.

\bibitem{yan22personalized}
An~Yan, Chaosheng Dong, Yan Gao, Jinmiao Fu, Tong Zhao, Yi~Sun, and Julian
  McAuley.
\newblock Personalized complementary product recommendation.
\newblock In {\em WWW}, 2022.

\bibitem{recbole[1.0]}
Wayne~Xin Zhao, Shanlei Mu, Yupeng Hou, Zihan Lin, Yushuo Chen, Xingyu Pan,
  Kaiyuan Li, Yujie Lu, Hui Wang, Changxin Tian, Yingqian Min, Zhichao Feng,
  Xinyan Fan, Xu~Chen, Pengfei Wang, Wendi Ji, Yaliang Li, Xiaoling Wang, and
  Ji{-}Rong Wen.
\newblock Recbole: Towards a unified, comprehensive and efficient framework for
  recommendation algorithms.
\newblock In {\em {CIKM}}, pages 4653--4664. {ACM}, 2021.

\bibitem{ziegler2005improving}
Cai-Nicolas Ziegler, Sean~M McNee, Joseph~A Konstan, and Georg Lausen.
\newblock Improving recommendation lists through topic diversification.
\newblock In {\em Proceedings of the 14th international conference on World
  Wide Web}, pages 22--32, 2005.

\end{thebibliography}

\end{document}